\newcommand{\AP}[3]{Ann.\ Phys.\ {\bf #1},\ #2 (#3)}
\newcommand{\NPA}[3]{Nucl.\ Phys.\ {\bf A#1},\ #2 (#3)}
\newcommand{\NPB}[3]{Nucl.\ Phys.\ {\bf B#1},\ #2 (#3)}

\newcommand{\PLB}[3]{Phys.\ Lett.\ B\ {\bf #1},\ #2 (#3)}
\newcommand{\PR}[3]{Phys.\ Rep.\ {\bf #1},\ #2 (#3)}
\newcommand{\PRL}[3]{Phys.\ Rev.\ Lett.\ {\bf #1},\ #2 (#3)}

\newcommand{\PRC}[3]{Phys.\ Rev.\ C\ {\bf #1},\ #2 (#3)}
\newcommand{\PRD}[3]{Phys.\ Rev.\ D\ {\bf #1},\ #2 (#3)}
\newcommand{\JPG}[3]{J.\ Phys.\ G\ {\bf #1},\ #2 (#3)}

\newcommand{\ZPC}[3]{Z.\ Phys.\ C\ {\bf #1},\ #2 (#3)}

\newcommand{\EPJA}[3]{Eur.\ Phys.\ J.\ A\ {\bf #1},\ #2 (#3)}
\newcommand{\PTP}[3]{Prog.\ Theo.\ Phys.\ {\bf #1},\ #2 (#3)}

\newcommand{\diracslash}[1]{#1\llap{/\kern2pt}}

\newcommand{\be}{\begin{equation}}
\newcommand{\ee}{\end{equation}}
\newcommand{\bea}{\begin{eqnarray}}
\newcommand{\eea}{\end{eqnarray}}
\newcommand{\ba}[1]{\begin{array}{#1}}
\newcommand{\ea}{\end{array}}

\documentclass[superscriptaddress,secnumarabic,amssymb,nobibnotes,nofootinbib,aps,prc,showpacs]{revtex4}
\usepackage{graphicx}

\begin{document}

\title{ Chiral symmetry breaking, color superconductivity
and color neutral quark matter: a variational approach}

\author{Amruta Mishra}
\email{mishra@th.physik.uni-frankfurt.de}
\affiliation{ Institute f\"ur Theoretische Physik, 
Universit\"at Frankfurt,
D-60054 Frankfurt, Germany}

\author{Hiranmaya Mishra}
\email{hm@prl.ernet.in}

\affiliation{Theory Division, Physical Research Laboratory,
Navrangpura, Ahmedabad 380 009, India}

\date{\today} 

\def\be{\begin{equation}}
\def\ee{\end{equation}}
\def\bearr{\begin{eqnarray}}
\def\eearr{\end{eqnarray}}
\def\zbf#1{{\bf {#1}}}
\def\bfm#1{\mbox{\boldmath $#1$}}
\def\hf{\frac{1}{2}}
\begin{abstract}
We investigate the vacuum realignment for chiral symmetry breaking and 
color superconductivity at finite density in  Nambu-Jona-Lasinio model in
a variational method. The treatment allows us to investigate 
simultaneous formation of condensates in quark antiquark as well as in
diquark channels.
The methodology involves an explicit construction of a variational
ground state and minimisation of the thermodynamic potential.
Color and electric charge neutrality conditions are imposed through
introduction of appropriate chemical potentials. Color and flavor dependent
condensate functions are determined through minimisation of the thermodynamic
potential. The equation of state is calculated. 
Simultaneous existence of a mass gap and superconducting gap 
is seen in a small window of quark chemical potential 
within the model when charge neutrality conditions are not imposed.
Enforcing color and electric charge neutrality conditions 
gives rise to existence of gapless superconducting
modes depending upon the magnitude of the gap and the difference of the
chemical potentials of the condensing quarks.
\end{abstract}

\pacs{12.38.Mh, 24.85.+p} 

\maketitle

 \section{Introduction}

 The structure of vacuum in
Quantum Chromodynamics (QCD) is one of the most interesting
questions in strong interaction physics \cite{sur}. The evidence for
quark and gluon condensates in vacuum is a reflection of its complex nature
\cite{svz}, whereas chiral symmetry breaking is an essential feature 
 in the description of the low mass hadron properties. Due to the 
nonperturbative nature of QCD in this regime different effective models
have been used to understand the nature of chiral symmetry breaking
\cite{chrl}. 
These have been constructed, for the most part, in the framework of 
a Nambu-Jona- Lasinio (NJL) model with a four fermion interaction. 

Recently there has been a lot of interest in strongly interacting 
matter at high densities.
In particular, a color superconducting phase for it involving
diquark condensates has been considered with a
gap of about 100 MeV . The studies have been done with an effective 
four fermion interaction between quarks \cite{wil}, direct instanton approach 
\cite{sursc} or a perturbative QCD calculation at finite density 
\cite{pisarski}. There has also been
a study of this phase in NJL model \cite{sarah}. Possibility of diquark
condensates alongwith quark antiquark condensates has been considered in Ref.
\cite{{berges},{npa},{mei},{blaschke},{kunihiro}}. The natural place to 
look for such a phase seems to be in the interior of compact stellar objects 
like neutron star.  However, to apply it to the case of neutron stars, 
the color and electric charge neutrality conditions need to be imposed for the 
bulk quark matter. Such an attempt has been made in Ref.\cite{krisch} as 
well as in Ref. \cite{reddy} where the lighter up and
down quarks form two flavor color superconducting (2SC) matter
while the strange quark do not participate in pairing. 
It has been shown, based upon comparison of
free energy that a two flavor color superconducting phase
would be absent in the core of
neutron stars \cite{krisch}. Within NJL model in Ref.\cite{reddy}
it has been argued that such conclusions are consistent except for a small
window in  density range where superconducting phase is 
possible. There have also been studies to include the possibility of mixed 
phases \cite{bubmix} of superconducting matter demanding neutral
matter on the average. In this context, there have been attempts to study
the implications of vector interactions on the structure of the phase diagram 
\cite{kunihiro}.

We had applied a different approach to study the problem in Ref.\cite{npa}.
We considered a variational approach with an explicit assumption for
the ground state having both quark antiquark and diquark condensates. The
actual calculations are carried out for the NJL model such that the
minimisation of the free energy density determines which condensate will
exist at what density. In the present work we generalise the approach
of Ref. \cite{npa} to include the conditions of color and electric 
charge neutrality. This leads to condensate functions which depend
upon both color and flavor. Although, for simplicity, we shall be
considering color superconductivity, this can be generalised to the three
flavor case to include color flavor locking. In fact here we shall also consider
three flavors but having the u and d quarks taking part in diquark condensation.
Although it might look rather complicated at the outset, the niceity of the
approach is that within the model one can solve for the condensate functions
explicitly  which are flavor and color dependent.

We organize the paper as follows. In the next section we discuss the ansatz
state with quark antiquark as well as diquark condensates. In section 3
we consider Nambu Jona-Lasinio model Hamiltonian and calculate the
expectation value with respect to the ansatz state to compute the thermodynamic 
potential. We minimise the thermodynamic potential to calculate
all the ansatz functions and the resulting mass as well as superconducting 
gap equations here. In section 4 we give the results and discuss them.
 Finally we summarise and conclude in section 5.
\section{ An ansatz for the ground state}
 As noted earlier we shall 
include here the effects of both chiral symmetry breaking as well as
diquark pairing. 
For the consideration of chiral symmetry breaking, we denote the
perturbative vacuum state with chiral symmetry as $|0\rangle$. 
We shall then assume a specific vacuum realignment which breaks chiral 
symmetry because of interaction.

Let us note first the quark field operator
expansion in momentum space given as
 \cite{hmnj,amspm}
\begin{eqnarray}
\psi (\zbf x )\equiv &&\frac{1}{(2\pi)^{3/2}}\int \tilde\psi(\zbf k)
e^{i\zbf k\cdot\zbf x}d\zbf k \nonumber\\ 
=&&\frac{1}{(2\pi)^{3/2}}\int \left[U_0(\zbf k)q^0_I(\zbf k )
+V_0(-\zbf k)\tilde q^0_I(-\zbf k )\right]e^{i\zbf k\cdot \zbf x}d \zbf k,
\label{psiexp}
\end{eqnarray}
where 
\begin{eqnarray}
U_0(\zbf k )=&&\left(\begin{array}{c}\cos(\frac{\phi^0}{2})\\
\zbf \sigma \cdot \hat k \sin(\frac{\phi^0}{2})
\end{array}\right),\;\;
V_0(-\zbf k )=
\left( \begin{array}{c} -\zbf \sigma \cdot \hat k  \sin(\frac{\phi^0}{2})
\\ \cos(\frac{\phi^0}{2})\end{array}
\right).
\label{uv0}
\end{eqnarray}
The superscript $0$ indicates that the operators 
$q_I^0$ and $\tilde q_I^0$ are
two component ones which annihilate or create quanta acting upon
the perturbative or the chiral vacuum $|0\rangle$. We have suppressed 
here the color and flavor indices of the quark field operators. The function 
$\phi^0(\zbf k)$ in the spinors in Eq.(\ref{uv0}) are given as 
$\cot{\phi_i^0}=m_i/|\zbf k|$,  for free massive fermion fields, 
$i$ being the flavor index. For massless fields
$\phi^0(|\zbf k|)=\pi/2$.

We now consider vacuum destabilisation leading to chiral symmetry breaking
\cite{npa,hmnj,amspm} described by,
\begin{equation} 
|vac\rangle={\cal U}_Q|0\rangle,
\label{u0}
\end{equation} 
where
\begin{equation}
{\cal U}_Q=\exp\left (
\int q_I^{0i}(\zbf k)^\dagger(\bfm {\sigma }\cdot\zbf k) h_i(\zbf k)
 \tilde q_I^{0i} (\zbf k)d\zbf k-h.c.\right ).
\label{u1}
\end{equation}
 
In the above, $h_i(\zbf k)$  is a
real function of $|\zbf k|$ which describes vacuum realignment for
quarks of a given flavor $i$. We shall take 
the condensate function $h(\zbf k)$ to be the same for u and d quarks and
$h_3(\zbf k)$ as the chiral condensate function for the s-quark.
Clearly, a nontrivial $h_i(\zbf k)$ shall break chiral
symmetry. Sum over three colors and three flavors is understood in the
exponent of ${\cal U}_Q$
in Eq.(\ref{u1}).

Having defined the state as in Eq.(\ref{u0}) for chiral symmetry breaking,
we shall next define the state involving diquarks. We note that
as per BCS result such a state will be dynamically favored
if there is an  attractive interaction between the quarks \cite{bcs}.
Such an interaction exists in QCD in the qq color antitriplet, Lorentz 
scalar and isospin singlet channel. In the flavor antisymmetric channel
 the interaction can be scalar, pseudoscalar or vector whereas in 
flavor symmetric channel only the axial vector channel is attractive.
In the present work, we shall consider the ansatz
 state involving diquarks as 
\begin{equation}
|\Omega\rangle={\cal U}_d|vac\rangle=\exp(B_d^\dagger-B_d)|vac\rangle,
\label{omg}
\end{equation}
where
\begin{equation}
{B}_d ^\dagger=\int \left[q_r^{ia}(\zbf k)^\dagger
r f^{ia}(\zbf k) q_{-r}^{jb}(-\zbf k)^\dagger
\epsilon_{ij}\epsilon_{3ab}
+\tilde q_r^{ia}(\zbf k)
r f_1^{ia}(\zbf k) \tilde q_{-r}^{jb}(-\zbf k)
\epsilon_{ij}\epsilon_{3ab}\right]
d\zbf k.
\label{bd}
\end{equation}
\noindent 
In the above, $i,j$ are flavor indices, $a,b$ are the
color indices and $r(=\pm 1/2) $ is the spin index. As noted earlier we 
shall  have u,d ($i$=1,2) quark condensation.
We have also introduced  here (color, flavor dependent) functions
 $f^{ia}(\zbf{k})$ and
$f^{ia}_1(\zbf k)$ respectively for the diquark and diantiquark
channels. As may be noted the state constructed in Eq.(\ref{omg}) is spin
singlet and is antisymmetric in color and flavor.
Clearly, by construction $f^{ia}(\zbf k)=f^{jb}(\zbf k)$ with $i\neq j$ and
$a\neq b$. The corresponding Bogoliubov
transformation for the operators is given by

\begin{eqnarray}
\left[
\begin{array}{c}q_{Ir}^{ia\prime}(\zbf k)\\ 
 q_{I-r}^{kc\prime}(-\zbf k)^\dagger\end{array} 
\right]& =&
\left[
\begin{array}{cc} \cos {f^{ia}(\zbf k)}  &
 -2 r \epsilon_{ik}\epsilon_{3ac}\sin f^{kc}(\zbf{k})\\
 2 r \epsilon_{ki}\epsilon_{3ca}\sin f^{ia}(\zbf{k}) & \cos{f^{kc}(\zbf k)}
  \end{array}
 \right]
\left[
\begin{array}{c}q_{Ir}^{ia}(\zbf k)\\ 
\tilde q_{I-r}^{kc}(-\zbf k)\end{array} 
\right].
\label{uquqpd}
\end{eqnarray}
In a similar manner one can write down the Bogoliubov transformation for 
the antiquark operators corresponding to $|\Omega\rangle$ basis.

Finally, to include the effect of temperature and density we next write down the 
state at finite temperature and density $|\Omega(\beta,\mu)\rangle$  taking
a thermal Bogoliubov transformation over the state $|\Omega\rangle$ 
using thermofield dynamics (TFD) as described in ref.s \cite{tfd,amph4}.
We then have,
\begin{equation} 
|\Omega(\beta,\mu)\rangle={\cal U}_{\beta,\mu}|\Omega\rangle={\cal U}_{\beta,\mu}
{\cal U}_d{\cal U}_Q |0\rangle.
\label{ubt}
\end{equation} 
where ${\cal U}_{\beta,\mu}$ is
\begin{equation}
{\cal U}_{\beta,\mu}=e^{{\cal B}^{\dagger}(\beta,\mu)-{\cal B}(\beta,\mu)},
\label{ubm}
\end{equation}
with, 
\begin{equation}
{\cal B}^\dagger(\beta,\mu)=\int \Big [
q_I^\prime (\zbf k)^\dagger \theta_-(\zbf k, \beta,\mu)
\underline q_I^{\prime} (\zbf k)^\dagger +
\tilde q_I^\prime (\zbf k) \theta_+(\zbf k, \beta,\mu)
\underline { \tilde q}_I^{\prime} (\zbf k)\Big ] d\zbf k.
\label{bth}
\end{equation}
In Eq.(\ref{bth}) the ansatz functions $\theta_{\pm}(\zbf k,\beta,\mu)$
will be related to quark and antiquark distributions and the underlined
operators are the operators in the extended Hilbert space associated with
thermal doubling in TFD method. In Eq.(\ref{bth}) we have suppressed
the color and flavor indices on the quarks as well as the functions
$\theta(\zbf k,\beta,\mu)$.
Note that we have a proliferation of functions in the ansatz
state $|\Omega,\beta,\mu\rangle$ -- the (flavor dependent) chiral condensate
 function, the (color flavor dependent) quark as well as antiquark condensate 
functions and the (color flavor dependent) thermal functions. All these 
functions are to be obtained by minimising the
thermodynamic potential. This will involve an 
assumption about the effective 
Hamiltonian. We shall carry out this minimisation
in the next section.

\section{Minimisation of thermodynamic potential and gap equations }

We shall work here in a Nambu-JonaLasinio model which is based on relativistic
fermions interacting through local current- current couplings assuming that 
gluonic degrees of freedom can be frozen into point like effective interactions 
between the quarks. The Hamiltonian is given as

\be
{\cal H}=\sum_{i,a}\psi^{ia \dagger}(-i\bfm \alpha \cdot \bfm \nabla
+\gamma^0 m_i )\psi^{ia}
+\frac{g^2}{2}J_\mu^aJ^{\mu a}.
\label{ham}
\ee
Here $J_\mu^a=\bar\psi\gamma_\mu T^a\psi$ and, $m_i$ is
 the current quark mass which we shall take to be 
nonzero only for the case of strange quarks ($i$=3). Of the two superscripts
on the quark operators the first index '$i$' refers to the flavor 
index and the second index, $a$ refers to the color index.
The point interaction produces short distance singularities and 
to regulate the integrals we shall restrict the phase space inside 
the sphere $|\zbf p|< \Lambda$ -- the ultraviolet cutoff in the NJL model.

We next write down the expectation values of various 
operators in the variational ansatz state given in Eq. (\ref{ubt}). 
Using the fact
that the state in Eq. (\ref{ubt}) arises from successive Bogoliubov
transformations one can calculate these expectation values.
These expressions would be used to calculate thermal expectation value of the
Hamiltonian to compute the thermodynamic potential. With $\tilde\psi(\zbf k)$as
defined in Eq.(\ref{psiexp}), we evaluate the expectation values
\begin{equation}
\langle \Omega(\beta,\mu)
 |\tilde\psi_\alpha^{ia}(\zbf k)\tilde\psi ^{jb}_\beta(\zbf k')^{\dagger}
|\Omega(\beta,\mu)\rangle
=\delta^{ij}\delta^{ab}
\Lambda_{+\alpha\beta}^{ia}(\zbf k,\beta,\mu)\delta(\zbf k-\zbf k'),
\label{psipsidb}
\end{equation}
and,
\begin{equation}
\langle \Omega(\beta,\mu)
|\tilde\psi_\beta(\zbf k)^{ia\dagger}\tilde\psi_\alpha^{jb}(\zbf k')
|\Omega(\beta,\mu)\rangle
=\delta^{ij}\delta^{ab}
\Lambda_{-\alpha\beta}^{ia,jb}(\zbf k,\beta,\mu)\delta(\zbf k-\zbf k'),
\label{psidpsib}
\end{equation}
where,
\begin{equation}
\Lambda_\pm^{ia}(\zbf k,\beta,\mu)
=\hf\left[1\pm \left( F_1^{ia}(\zbf k)
-F^{ia}(\zbf k)\right)\pm \big(\gamma^0\cos\phi^i (\zbf k)\big)
+\bfm\alpha\cdot\hat\zbf k\sin \phi^i
(\zbf k)\big)\big(1-F^{ia}(\zbf k)-F_1^{ia}(\zbf k)\big)
\right].
\label{prpb}
\end{equation}
Here, the effect of diquark condensates and their temperature and/or density
dependences are encoded in the functions $F^{ia}(\zbf k)$ and $F_1^{ia}
(\zbf k)$ given as
\begin{equation}
F^{ia}(\zbf k)=\sin^2\theta^{ia}_-(\zbf k)+\sin^2 f^{ia}\left(C_-^{ia}(\zbf k)-
\sin^2\theta_-^{ia}(\zbf k)\right)\left(1-\delta^{a3}\right),
\label{fkb}
\end{equation}
and,
\begin{equation}
F_1^{ia}(\zbf k)=\sin^2\theta^{ia}_+(\zbf k)+\sin^2 f_1^{ia}
\left(\cos^2\theta_+^{ia}(\zbf k)- S_+^{ia}(\zbf k)\right)
\left(1-\delta^{a3}\right).
\label{f1kb}
\end{equation}
Here, we have defined $C_-^{ia}=
|\epsilon^{ii'}\epsilon^{aa'}|\cos^2\theta_-^{i'a'}$ and
$S_+^{ia}=|\epsilon^{ii'}\epsilon^{aa'}|\sin^2\theta_+^{i'a'}$.
The $\delta^{a3}$ term indicates that the third color does not take part in 
diquark condensation. Further, we have introduced the  notation
$\phi_i(\zbf k)=\phi_i^0(\zbf k)-2 h_i(\zbf k)$.

\noindent
We also have

\bearr
\langle \Omega(\beta,\mu)| \psi^{ia}_\alpha(\zbf x)\psi^{jb}_\gamma
(\zbf 0)
|\Omega(\beta,\mu) \rangle
&=&-\frac{1}{(2\pi)^3}
\int e^{i\zbf k\cdot\zbf x}
{\cal {P}}_{+\gamma\alpha}^{ia,jb}(\zbf k,\beta,\mu)d\zbf k,\nonumber\\
\langle \Omega(\beta,\mu)| \psi^{ia\dagger}_\alpha(\zbf x)
\psi^{jb\dagger}_\gamma
(\zbf 0)
|\Omega(\beta,\mu) \rangle
&=&-\frac{1}{(2\pi)^3}\int e^{i\zbf k\cdot\zbf x}
{\cal {P}}_{-\alpha\gamma}^{ia,jb}(\zbf k,\beta,\mu)d\zbf k,
\label{psi}
\eearr

where, 
\bearr
{\cal{P}}_+^{ia,jb}(\zbf k,\beta,\mu)
&=&\frac{\epsilon^{ij}\epsilon^{3ab}}{4}\bigg[S^{ia,jb}(\zbf k)
\cos\left(\frac{\phi_i-\phi_j}{2}\right)\nonumber\\
&+&\left(\gamma^0
\cos \left(\frac{\phi_{i}+\phi_j}{2}\right)-\bfm\alpha\cdot
\hat\zbf k\sin\left(\frac{\phi_i+\phi_j}{2}\right)\right)A^{ia,jb}(\zbf k)
\bigg]\gamma_5 C,
\label{calpp}
\eearr
and,
\bearr
{\cal{P}}_-^{ia,jb}(\zbf k,\beta,\mu)
&=&\frac{\epsilon^{ij}\epsilon^{3ab}
C\gamma_5}{4}\bigg [S^{ia,jb}(\zbf k)
\cos\left(\frac{\phi_i-\phi_j}{2}\right)\nonumber\\
&+&
\left(\gamma^0\cos\left(\frac{\phi_i+\phi_j}{2}\right)
-\bfm\alpha\cdot \hat\zbf k\sin \left(\frac{\phi_i+\phi_j}{2}\right)
\right)A^{ia,jb}(\zbf k)
\bigg].
\label{calpm}
\eearr

\noindent Here, $C=i\gamma^2 \gamma^0$ is the charge conjugation matrix (we
use the notation of Bjorken and Drell) and the functions $S(\zbf k)$ and 
$A(\zbf k)$ are given as,
\be
S^{ij,ab}(\zbf k)=\sin\!2f^{ia}(\zbf k)\cos 2\theta^{ia,jb}_-(\zbf k,\beta,\mu)
+\sin\!2f_1^{ia}(\zbf k)\cos 2\theta^{ia,jb}_+(\zbf k,\beta,\mu),
\label{sk}
\ee
and,
\begin{equation}
A^{ij,ab}(\zbf k)=\sin\!2f^{ia}(\zbf k)\cos 2\theta^{ia,jb}_-(\zbf k,\beta,\mu)
-\sin\!2f_1^{ia}(\zbf k)\cos 2\theta_+^{ia,jb}(\zbf k,\beta,\mu),
\label{ak}
\end{equation}

\noindent In the above we have defined $\cos 2\theta^{ia,jb}_\pm
=1-\sin^2\theta_\pm^{ia}- \sin^2\theta_\pm^{jb}$, with ${i,j=1,2}$ 
being the flavor indices and $a,b=1,2$ being the color indices  and
 $i\neq j$, $a\neq b$.

Using Eq. (\ref{psidpsib}) we have for the kinetic energy of the light quarks
\bearr
T & \equiv &
 \langle \Omega(\beta,\mu)|
\psi^\dagger
(-i\bfm \alpha \cdot \bfm \nabla )\psi
| \Omega(\beta,\mu)\rangle \nonumber \\
 & = & \frac{2}{(2\pi)^3}\sum_{i=1,2,a=1,3}
\int d \zbf k |\zbf k|\left(1-\cos 2h_i(\zbf k)(1-F^{ia}-F_1^{ia})\right),
\label{tren}
\eearr
where, $F^{ia}$ and $F_1^{ia}$ are defined in equations (\ref{fkb})
and Eq.(\ref{f1kb}). We have also subtracted the vacuum contributions.

Similarly the contribution from the interaction term in Eq.(\ref{ham})
after subtracting out the zero point perturbative energy turns out to be
\begin{equation}
{V}\equiv
 \langle \Omega(\beta,\mu)|\frac{g^2}{2}J_\mu^aJ^{\mu a}
| \Omega(\beta,\mu)\rangle
=V_1+ V_2
\label{v}
\ee
Here, the contribution $V_1$ arises from contracting 
$\psi$ with a $\psi^\dagger$using Eq.s (\ref{psipsidb}),
(\ref{psidpsib})
and is given as
\be
V_1=\frac{g^2 }{2}\sum_{i=1,2}\left(\sum_{a=1,3}{I_v^{ia}}^2
-2\sum_{a=1,3}{I_s^{ia}}^2\right),
\label{v1}
\ee
with,
\be
I_v^{ia}=\frac{1}{(2\pi)^3}\int d \zbf k(F^{ia}-F_1^{ia})
\label{i1}
\ee
and
\be
I_s^{ia}=\frac{1}{(2\pi)^3}\int d \zbf k(1-F^{ia}-F_1^{ia})\sin 2h_i(\zbf k).
\label{i2}
\ee
In this expression, we have neglected terms of the order $1/N_c^2$ 
compared to unity.
The term $V_2$ arises from contracting $\psi$ and a $\psi$ and $\psi^\dagger$
with another $\psi^\dagger$ using Eq. (\ref{calpp})and (\ref{calpm})
and we have
\be
V_2=-\frac{4}{3}g^2 I_3^{11,22}I_3^{12,21}
\label{v2}
\ee
with
\be
I_3^{ia,jb}=\frac{1}{(2\pi)^3}\int d \zbf k S^{ia,jb}(\zbf k)
\cos\left(\frac{\phi_i-\phi_j}{2}\right)
\label{i3}
\ee

\noindent where, $S^{ia,jb}$ has been defined in Eq.(\ref{sk}). 

To calculate the thermodynamic potential we shall have to specify the
chemical potentials relevant for the system.
Here we shall be interested in the form of quark matter that might be present
in compact stars older than few minutes so that 
chemical equilibriation under weak
interaction is there. The relevant chemical potentials in this case then
are the baryon chemical potential $\mu_B=3\mu$, the chemical potential 
$\mu_E$ associated with electromagnetic charge $Q=diag(2/3,-1/3,-1/3)$
in flavor space, 
and the two color electrostatic chemical potentials $\mu_3$ and $\mu_8$ 
corresponding to $U(1)_3\times U(1)_8$ subgroup of the color gauge symmetry
generated by cartan subalgebra $Q_3=diag(1/2,-1/2,0)$ and 
$Q_8=diag(1/3,1/3,-2/3)$ in the color space. Thus the chemical potential
is a diagonal matrix in color and flavor space, and is given by
\be
\mu_{ij,ab}=(\mu\delta_{ij}+Q_{ij}\mu_E)\delta_{ab}
+(Q_{3ab}+Q_{8ab}\mu_8)\delta_{ij}.
\label{muij}
\ee
Here, $i,j$ are flavor indices and $a,b$ are color indices.

The total thermodynamic potential, including the contribution from the
electrons, is then given by
\be
\Omega=T+V-\langle \mu N\rangle-\frac{1}{\beta}s+\Omega_e
\ee
where, we have denoted
\be
\langle \mu N\rangle=\langle \psi^{ia\dagger}\mu_{ij,ab}\psi^{jb}\rangle
=2\sum_{i,a}\mu^{ia}I_v^{ia}
\ee
with $\mu^{ia} $ being the chemical potential for the quark of flavor $i$
and color $a$, which can be expressed in terms of the chemical potentials
$\mu$, $\mu_E$,$\mu_3$ and $\mu_8$ using Eq.(\ref{muij}). $\Omega_e=
-\mu_E^4/12\pi^2$ is the electron free energy.

Finally, for the entropy density for the quarks we have \cite{tfd}
\bearr
s & = & -\frac{2}{(2\pi)^3}\sum_{i,a}\int d \zbf k
\Big ( \sin^2\theta^{ia}_-\ln \sin^2\theta^{ia}_-
+\cos^2\theta^{ia}_-\ln \cos^2\theta^{ia}_- \nonumber \\ 
& + & \sin^2\theta^{ia}_+ln \sin^2\theta^{ia}_+
+\cos^2\theta^{ia}_+ln \cos^2\theta^{ia}_+\Big ).
\label{ent}
\eearr

Now if we minimise the thermodynamic potential $\Omega$ with respect to
$h _i (\zbf k)$, we get
\be
\tan 2 h_i(\zbf k)
= \frac{ (M_i -m_i) k}{\epsilon_i^2 + g^2 \sum_{a=1,3}I_s^{ia} m_i}
\label{tan2h}
\ee
\noindent where, $M_i= m_i + g^2 \sum_b I^{ib}_s$ and $\epsilon_i
=\sqrt{(\zbf k^2 +m_i^2)}$.
Substituting this back in Eq.(\ref{i2}) we have the mass gap equation
for the light quarks ($m_i$=0) 
\be
M_j= g^2\sum_aI_s^{ja}=\frac{ g^2}{(2\pi)^3}
\int \frac{M_j}{\sqrt{\zbf k^2 +M_j^2}} \sum_{a=1,3}(1-F^{ja}-F^{ja}_1)
 d \zbf k.
\label{mgap}
\ee
Clearly, the above includes the effect of diquark condensates 
as well as temperature and density through the functions $F$ and $F_1$ 
given in Eq.s (\ref{fkb})
and (\ref{f1kb}) respectively.

Next, minimising the thermodynamic potential with respect to the diquark 
condensate functions leads, to 
\be
\tan 2f^{12}(\zbf k)=\frac{\Delta_{12}}{\bar \epsilon-\bar \nu_{12}}
\cos(\frac{\phi_1-\phi_2}{2})
\label{tan2f}
\ee
\be
\tan 2f^{11}(\zbf k)=\frac{\Delta_{11}}{\bar \epsilon-\bar \nu_{11}}
\cos(\frac{\phi_1-\phi_2}{2})
\label{tan2f11}
\ee
 In the above $\bar\epsilon=(\epsilon_1+\epsilon_2)/2$, 
$\bar\nu_{11}= (\nu_{11}+\nu_{22})/2$,
$\bar\nu_{12}= (\nu_{12}+\nu_{21})/2$.
$\nu^{ia}$ is the {\em interacting} chemical potential
given as
\be
\nu^{ia}=\mu^{ia}-\frac{g^2}{4}\rho^i
\label{nui}
\ee
which may be expected in presence of  a vectorial current--current interaction .
In the above, $\rho^i=2 \sum _ {a=1,3}I_v^{ia}$, with $I_v^{ia}$ as defined in
equantion (\ref{i1}).
Thus, it may be noted that the diquark condensate functions depend upon
the {\em average} energy and the {\em average} chemical potential
of the quarks that condense. We also note here that the 
diquark condensate functions depends upon the masses of the two quarks which
condense through the function $\cos \big ((\phi_1-\phi_2)/2\big )$ with 
$\cos\phi_i=\sin 2 h_i=M_i/\epsilon_i$, for u,d quarks which could be different
when charge neutrality condition is imposed. Such a normalisation
factor is always there when
the condensing fermions have different masses as has been noted in Ref.
\cite{aichlin} in the context of CFL phase. 

In an identical manner the di-antiquark condensate functions are 
calculated to be
\be
\tan 2f_1^{12}(\zbf k)=\frac{\Delta_{12}}{\bar \epsilon+\bar \nu_{12}}
\cos(\frac{\phi_1-\phi_2}{2})
\label{tan2f1}
\ee
\be
\tan 2f^{11}_1(\zbf k)=\frac{\Delta_{11}}{\bar \epsilon+\bar \nu_{11}}
\cos(\frac{\phi_1-\phi_2}{2})
\label{tan2f111}
\ee
Further, in Eq.s (\ref{tan2f}),(\ref{tan2f11}),(\ref{tan2f1}),(\ref{tan2f111}),
$\Delta_{12}=(2g^2/3) I_3^{11,22}$, $\Delta_{11}=(2g^2/3) I_3^{12,21}$
which satify the equations
\bearr
\Delta_{12}=\frac{2g^2}{3(2\pi)^3}&\int & d\zbf k 
\bigg[\frac{\Delta_{11}}{\sqrt{{\bar\xi_{-12}}^2+\Delta_{11}^2\cos^2
\left(\frac{\phi_1-\phi_2}{2}\right)}}
\left(\cos^2\theta_-^{11}-\sin^2\theta_-^{22}\right)\nonumber\\
&+&
\frac{\Delta_{11}}{\sqrt{{\bar\xi_{+12}}^2+\Delta_{11}^2\cos^2
\left(\frac{\phi_1-\phi_2}{2}\right)}}
\left(\cos^2\theta_+^{11}-\sin^2\theta_+^{22}\right)\bigg ]
\cos \left(\frac{\phi_1-\phi_2}{2}\right)
\label{del12}
\eearr
and,
\bearr
\Delta_{11}=\frac{2g^2}{3(2\pi)^3}&\int &d\zbf k 
\bigg[\frac{\Delta_{12}}{\sqrt{{\bar\xi_{-11}}^2+\Delta_{12}^2\cos^2
\left(\frac{\phi_1-\phi_2}{2}\right)}}
\left(\cos^2\theta_-^{11}-\sin^2\theta_-^{22}\right)\nonumber\\
&+&\frac{\Delta_{12}}{\sqrt{{\bar\xi_{+11}}^2+\Delta_{12}^2\cos^2
\left(\frac{\phi_1-\phi_2}{2}\right)}}
\left(\cos^2\theta_+^{11}-\sin^2\theta_+^{22}\right)\bigg].
\cos \left(\frac{\phi_1-\phi_2}{2}\right)
\label{del11}
\eearr
In the above, 
${\bar \xi }_{\pm ia}=\bar \epsilon \pm {\bar \nu}^{ia}$. 
Finally, minimisation of the thermodynamic potential with respect to the
thermal functions $\theta_{\pm}(\zbf k)$ gives
\be
\sin^2\theta_\pm^{ia}=\frac{1}{\exp(\beta\omega_\pm^{ia})+1}
\label{them}
\ee

 \noindent Various $\omega^{ia}$s are given as follows.
$$\omega_-^{11} =\omega_- +\delta_\epsilon-\delta_\nu^{11}$$
$$\omega_-^{12} =\omega_- +\delta_\epsilon-\delta_\nu^{12}$$
$$\omega_-^{21} =\omega_--\delta_\epsilon+\delta_\nu^{12}$$
$$\omega_-^{22} =\omega_--\delta_\epsilon+\delta_\nu^{11}$$
$$\omega_+^{11} =\omega_++\delta_\epsilon+\delta_\nu^{11}$$
$$\omega_+^{12} =\omega_++\delta_\epsilon+\delta_\nu^{12}$$
$$\omega_+^{21} =\omega_+-\delta_\epsilon-\delta_\nu^{11}$$
\be
\omega_+^{22} =\omega_+-\delta_\epsilon-\delta_\nu^{11}
\label{disps}
\ee

\noindent and, finally,
for the noncondensing colors $\omega_{\pm}^{i3}=\epsilon^i{\pm}\nu^{i3}$.
As already mentioned, the first index refers to flavor and the second 
index refers to color. Here $\omega_\pm=\sqrt{\Delta^2\cos^2((\phi_1-\phi_2)/2)
+\bar\xi_{\pm}^2}$,
$\delta _ \epsilon=(\epsilon_1-\epsilon_2)/2$ is half the energy difference
of the two quarks which condense and e.g. $\delta_\nu^{11}=
(\nu_{11}-\nu_{22})/2$, is half the difference of the chemical potentials
of the two quarks which condense. Note that in the absence of imposing the
charge neutrality condition all the four quasi particles will have the
same energy $\omega_-$. Thus it is possible to have zero modes when charge
neutrality condition is imposed depending upon the values of $\delta_\epsilon$
and $\delta_\nu$. So, although we shall have nonzero order parameter $\Delta$,
there will be fermionic zero modes or the gapless superconducting phase \cite
{abrikosov, krischprl}. We shall discuss more about it section 4.

Next, let us focus our attention for the specific case of 
superconducting phase and the chemical potential associated with it. First
let us note that the diquark condensate functions depend upon the 
average of the chemical potentials of the quarks that condense. Since
this is independent of $\mu_3$ we can choose $\mu_3$ to be zero.
In that case, $\mu^{11}=\mu +2/3 \mu_E+\mu_8/\sqrt{3}=\mu^{12}$
and also $\mu^{21}=\mu -1/3 \mu_E+\mu_8/\sqrt{3}=\mu^{22}$.
This also means that the average chemical potential of both
the condensing quarks are the same and is equal to 
$\bar\mu=\mu+1/6\mu_E+\mu_8/\sqrt{3}$.
Further ${\delta_\nu}^{12}=\mu_E/2-(g^2/4)(\rho_1-\rho_2)/2={\delta_\nu}^{11}
\equiv\delta_\nu$. 
In such a situation, Eq.(\ref{del12}) and Eq.(\ref{del11}) 
are both identical and hence we shall have only one 
superconducting gap equation given as

\be
\Delta=\frac{2g^2}{3(2\pi)^3}\int d \zbf k
\left[\frac{\Delta}{\omega_-}
\left(\cos^2\theta_-^{12}-\sin^2\theta_-^{21}\right)+
\frac{\Delta}{\omega_+}
\left(\cos^2\theta_+^{12}-\sin^2\theta_+^{21}\right)\right]
\cos\left(\frac{\phi_1-\phi_2}{2}\right)
\label{scgap}
\ee

With this condition,it is clear from Eq.s(\ref{disps})
that the quasi particle energies  for each flavor becomes degenerate 
for both the colors which take part in condensation. Thus the quasi
particle energies now become $\omega_1=\omega_-+\delta$
and $\omega_2=\omega_--\delta$, 
for u and d quarks respectively, 
with $\delta=\delta_\epsilon-\delta_\nu$.

Now making the use of this dispersion relations and the mass
gap equations Eq.(\ref{mgap}) and the superconducting gap
equation Eq.(\ref{scgap}), the thermodynamic potential at zero 
temperature becomes, with $\delta=\delta_\epsilon-\delta_\nu$,
\bearr
\Omega_{u,d}&=& \frac{8}{(2\pi)^3}\int d\zbf k\left[|\zbf k|-
\frac{1}{2}(\omega_-+\omega_+)\right]\nonumber\\
&+& \frac{4}{(2\pi)^3}\int d\zbf k \left[(\omega_-+\delta)\theta(-\omega_1)
+(\omega_--\delta) \theta(-\omega_2)\right]\nonumber\\
&+& \frac{3\Delta^2}{g^2}-\frac{g^2}{8}\sum_{i=1,2}\rho_i^2+\frac{1}{g^2}
(M_1^2+M_2^2)\nonumber\\
&-& \sum_{i=1,2}\nu^{i3}\rho^{i3}+\frac{4}{(2\pi)^3}\int d^3 k 
\left[|\zbf k|
-\frac{\epsilon_1}{2}\theta(k-k_f^{13})-\frac{\epsilon_2}{2}
\theta( | \zbf  k|-k_f^{23})\right].
\label{omgud}
\eearr
The first three lines in Eq.(\ref{omgud}) correspond to the contribution
from the quarks taking part in the condensation while the last line
is the contribution from the third color for the two light quarks.

In an identical manner one can calculate the thermodynamic 
potential for the strange quark sector with strange quarks and anti-quarks 
for the vacuum structure. The contribution to the thermodynamic 
potential then is given by
\bearr
\Omega_s& =&\frac{2}{(2\pi)^3}\int d\zbf k\sum_{a=1,3}\left[
\sqrt{|\vec k|^2+m_s^2}-
\sqrt{\zbf k^2+M_3^2}\theta(|\zbf k|-k_f^{3a})\right]\nonumber\\
&+&\frac{(M_3-m_s)^2}{g^2}-\frac{g^2}{8}\rho_3^2-\sum_a\nu^{3a}\rho^{3a}.
\label{omgs}
\eearr

\noindent Here, $\rho_{i}=\sum_{a=1,3}\rho^{ia}$ and $\rho^{ia}=2 I_v^{ia}$ for
$i,a=1,2$, with $I_v^{ia}$ given in Eq.(\ref{i1}). For the third color,
$\rho^{i3}= {(k_f^{i3})^3}/{3\pi^2}$ and for the strange quark,
$\rho^{3a}={(k_f^{3a})^3}/{3\pi^2}$.
The fermi momenta are given by the usual relation $k_f^{ia}=
(\nu_{ia}^2-M_i^2)^{1/2}$.

The total thermodynamic potential is given by  including the 
contribution from the electrons and is given by
\be
\Omega=\Omega_{u,d}+\Omega_{s}-\frac{\mu_E^4}{12\pi^2}.
\label{thpot}
\ee
Thus the thermodynamic potential is a function of four parameters: the three
mass gaps and a superconducting gap which needs to be minimised
subjected to the conditions of electrical and color neutrality. 
The electric and charge neutrality constraints are given respectively as
\be
Q_E=\frac{2}{3}\rho_1-\frac{1}{3}\rho_2-\frac{1}{3}{\rho_3}
-\rho_e=0,
\label{qe}
\ee
and,
\be
Q_8 =\frac{1}{\sqrt{3}}\sum_ i (\rho^{i1} + \rho ^{i2}
- 2 \rho ^{i3}) =0.
\label{q8}
\ee

\noindent Eq.(\ref{thpot}--\ref{q8}) and the superconducting 
gap equation Eq.(\ref{scgap})  constitute the
basis of the numerical calculations that we discuss below.

\section{Results and discussions}

For numerical calculations we have taken the values of
the parameters of NJL model as follows: $g^2\Lambda^2=17.6$,
$\Lambda=0.68$ GeV as typical values giving reasonable
vacuum properties \cite{mei,aichlin,bubnp}. We might note here
that the coupling $g^2$ introduced here is related to the 
usual scalar coupling of NJL model e.g. in Ref.\cite{bubnp} $G$ 
as $g^2=8 G$.

Current quark masses for u and d quarks are
taken as zero and the current quark mass for strange quark is taken as 
0.12 GeV. With this choice of parameters, the constituent quark 
masses at zero temperature and density are given as $M_1=0.35$ GeV=$M_2$,
and for strange quark $M_3=0.575$ GeV. 

In Fig.\ref{figmass0}, we have plotted the variation of masses with 
baryon density without diquark condensation and without imposition 
of the charge neutrality conditions. 
\noindent The decrease of masses with density is a reflection of
the decrease of quark condensates with density.

\begin{figure}
\includegraphics[width=8cm,height=8cm]{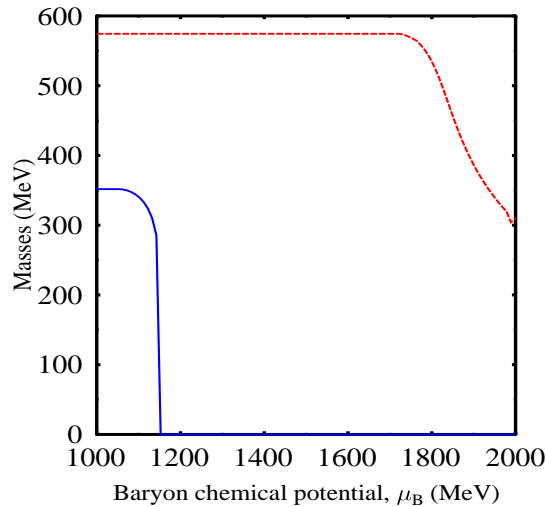}
 \caption{\label{figmass0} Masses of the quarks as a function of baryon chemical potential. 
 Solid curve refers to masses of u and d quarks while the dashed
 curve refers to strange quark mass.}
 \end{figure}

Let us next discuss the case with diquark condensates along with quark 
antiquark condensates without imposing the charge neutrality condition.
The numerical calculation for this case proceeds as follows.
For a given chemical baryon potential,(and with $\mu_3$=0= $\mu_8$), the 
thermodynamic potential Eq.(\ref{thpot}) is minimised with respect to
the quark masses subjected to self consistently determining the interacting 
chemical potential using Eq.(\ref{nui}) and solving the superconducting 
gap equation Eq.(\ref{scgap}). In Fig. 2 we have plotted the masses and the 
superconducting gap as a function of baryon chemical potential. 
The behaviour of masses are almost same to that without the 
diquark condensates.
Here, however, we observe that for a small 
window in the quark chemical potential ( about
23 MeV), we have simultaneous existence of chiral symmetry breaking 
and color superconducting phase with the gap reaching a maximum upto 65 MeV
in this window.  Including vector interactions similar conclusion of 
simultaneous existence of both the condenstes was noted earlier in 
Ref.\cite{kunihiro}. However, in Ref.\cite{kunihiro} the vector interaction 
did not contribute to the superconducting gap equation nor to the mass gap
equation unlike the case here.
We have compared the pressure in both the cases in this regime and
it appears that in this small window, existence of both the condensates has 
higher pressure than having only the quark antiquark condensates in the
ground state.
Beyond this window  chiral symmetry is restored and the gap increases
monotonically with chemical potential till the effect of cut off is felt
and then it decreases. 
 \begin{figure}[htbp]
 \begin{center}
 \includegraphics[width=8cm,height=9cm]
{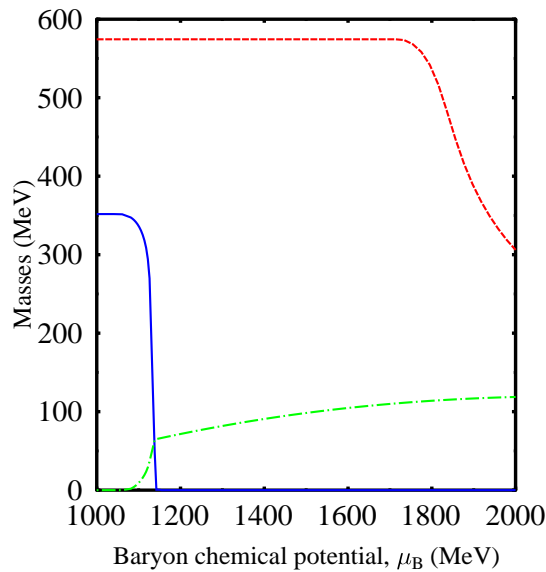}
 \end{center}
 \caption{\em Masses of the quarks and the superconducting gap
as functions of baryon chemical potential for the case $\mu_E$ =0=$\mu_8$.
 Solid, dashed and dot-dashed  curves  refer 
respectively to masses of up (down) quarks,
 strange quark and to
the superconducting gap.}
 \label{figdelta0}
 \end{figure}

       We next discuss the results when charge neutrality conditions
are imposed. As earlier, let us focus our attention first to the case without
the diquark condensates. The results are shown in Fig. 3. Here, 
the d- quark mass vanishes earlier than that of u quarks as the 
baryon chemical potential is increased. The reason is that to maintain
electrical charge neutrality conditions the d quark densities are
almost twice that of u quarks which makes the d quark antiquark 
condensates to vanish. In contrast to the rather sharp fall of the masses
as compared to without imposition of electrical neutrality conditions, the u 
quarks remain massive much after d quark become massless (about 80 MeV
in the quark chemical potential window). The magnitude of electric chemical 
potential increases with densities till strange quarks begin to appear 
beyond which it starts decreasing so that charge neutrality conditions are
maintained.
 \begin{figure}[htbp]
 \begin{center}
 \includegraphics[width=8cm,height=8cm]{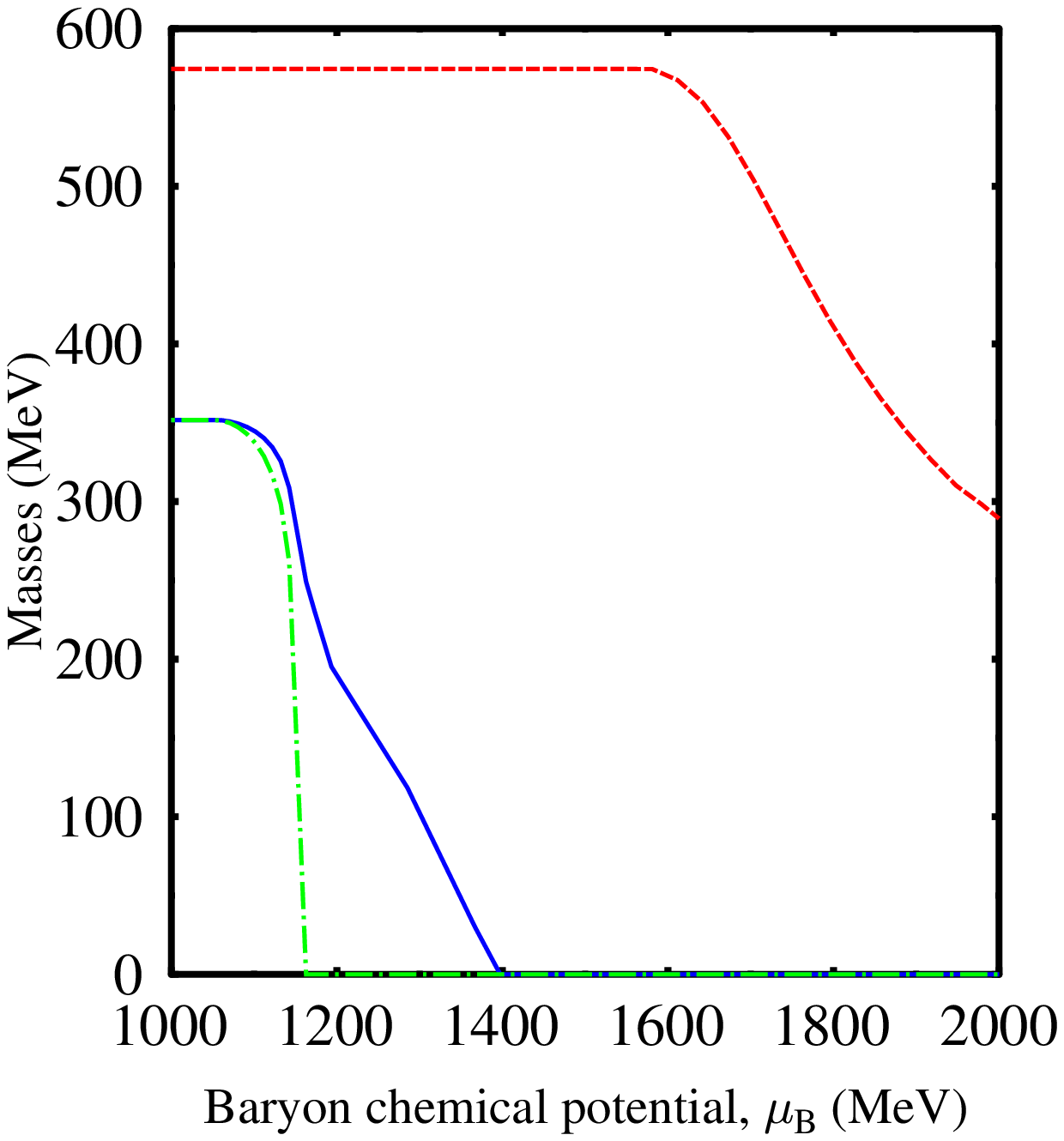}
 \end{center}
 \caption{\em Masses of the quarks without superconducting phase
when electrical charge neutrality condition is imposed.
 Solid, dot-dashed and dashed  curves  refer 
respectively to masses of up, down and strange quark mass. 
}
 \label{figmass}
 \end{figure}

\begin{figure}[htbp]
\begin{center}
 \includegraphics[width=8cm,height=8cm]{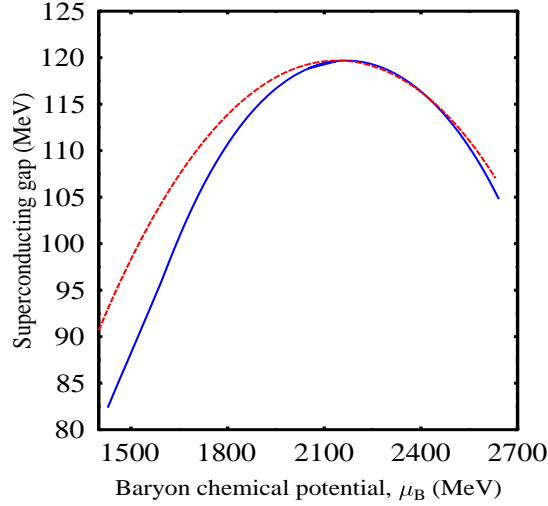}
\end{center}
\caption{\em Superconducting gap when color and electrical charge neutrality
condition is imposed (solid curve). The dashed curve corresponds to
when this condition is not imposed (i.e.$\mu_E$=0).}
\label{figdeln}
 \end{figure}

\begin{figure}[htbp]
\begin{center}
 \includegraphics[width=8cm,height=8cm]{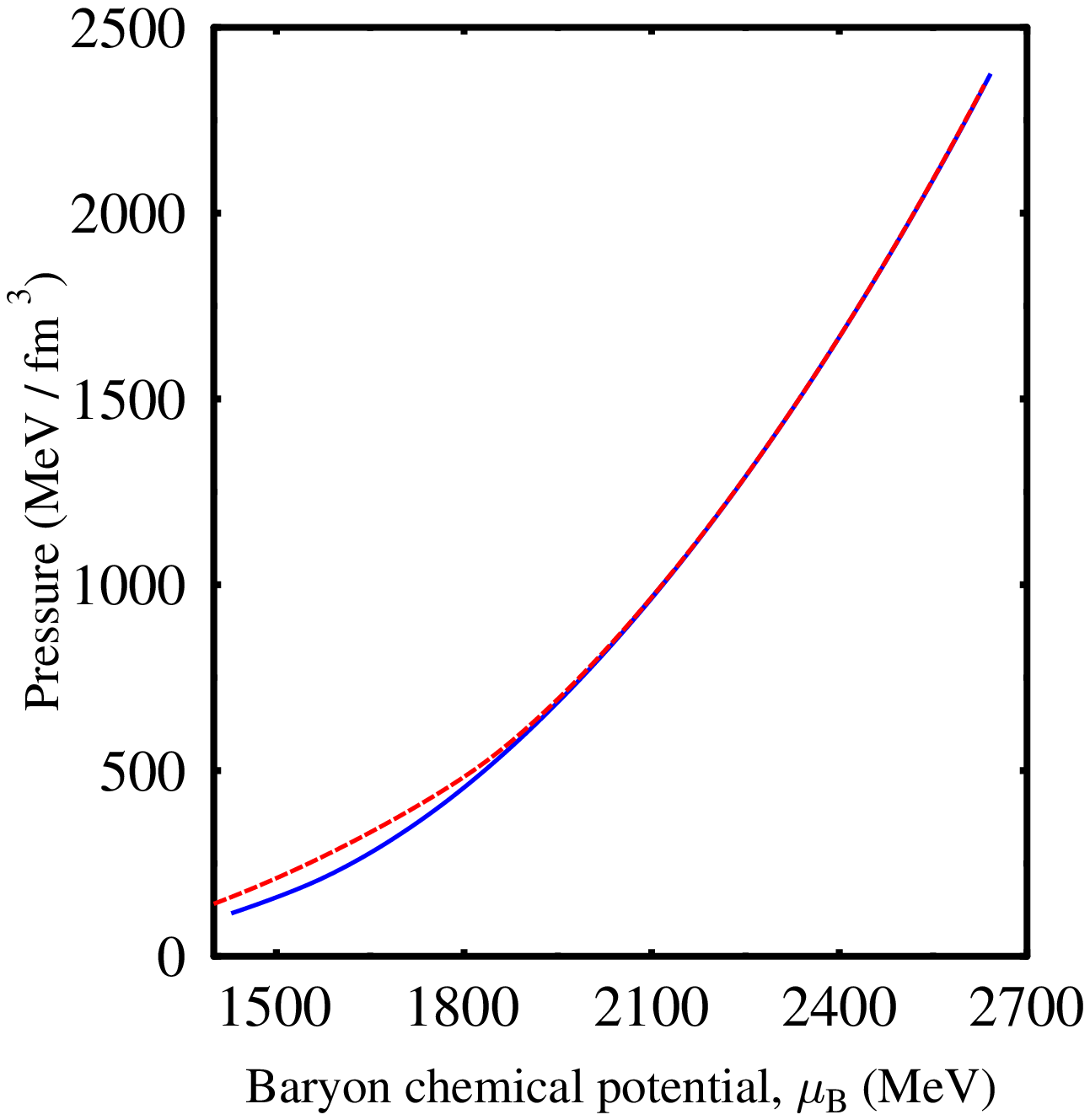}
\end{center}
\caption{\em Pressure as a function of baryon chemical potential
when color and electrical charge neutrality
condition is imposed (solid curve). The dashed curve corresponds to
when this condition is not imposed.}
\label{figpre}
 \end{figure}

Next we show the results with diquark condensates with neutrality conditions.
For a given quark chemical potential $\mu$,
the interacting chemical potentials are determined using Eq.(\ref {nui})
with a trial value of $\mu_E$ and $\mu_8$.
For high baryon chemical potential when chiral symmetry is restored for 
light quarks, the thermodynamic potential is varied with respect to the
strange quark constituent mass after solving the superconducting gap equation
Eq.(\ref{scgap}). The values of $\mu_E$ and $\mu_8$ are varied so that the
charge neutrality conditions Eq.(\ref{qe}) and Eq.(\ref{q8}) are satisfied.
The results are shown in Fig.4. The behaviour of the superconducting gap
is similar to that when charge neutrality condition are not imposed 
except that the magnitude of the gap decreases. The pressure
of color and electrical neutral superconducting phase is lower than
when neutrality conditions are not imposed as shown in Fig.\ref{figpre}.
Further, the pressure of color and electric 
neutral normal quark matter is always lower than that of the 
color and electrically neutral superconducting matter.

\begin{figure}[htbp]
\begin{center}
 \includegraphics[width=8cm,height=8cm]{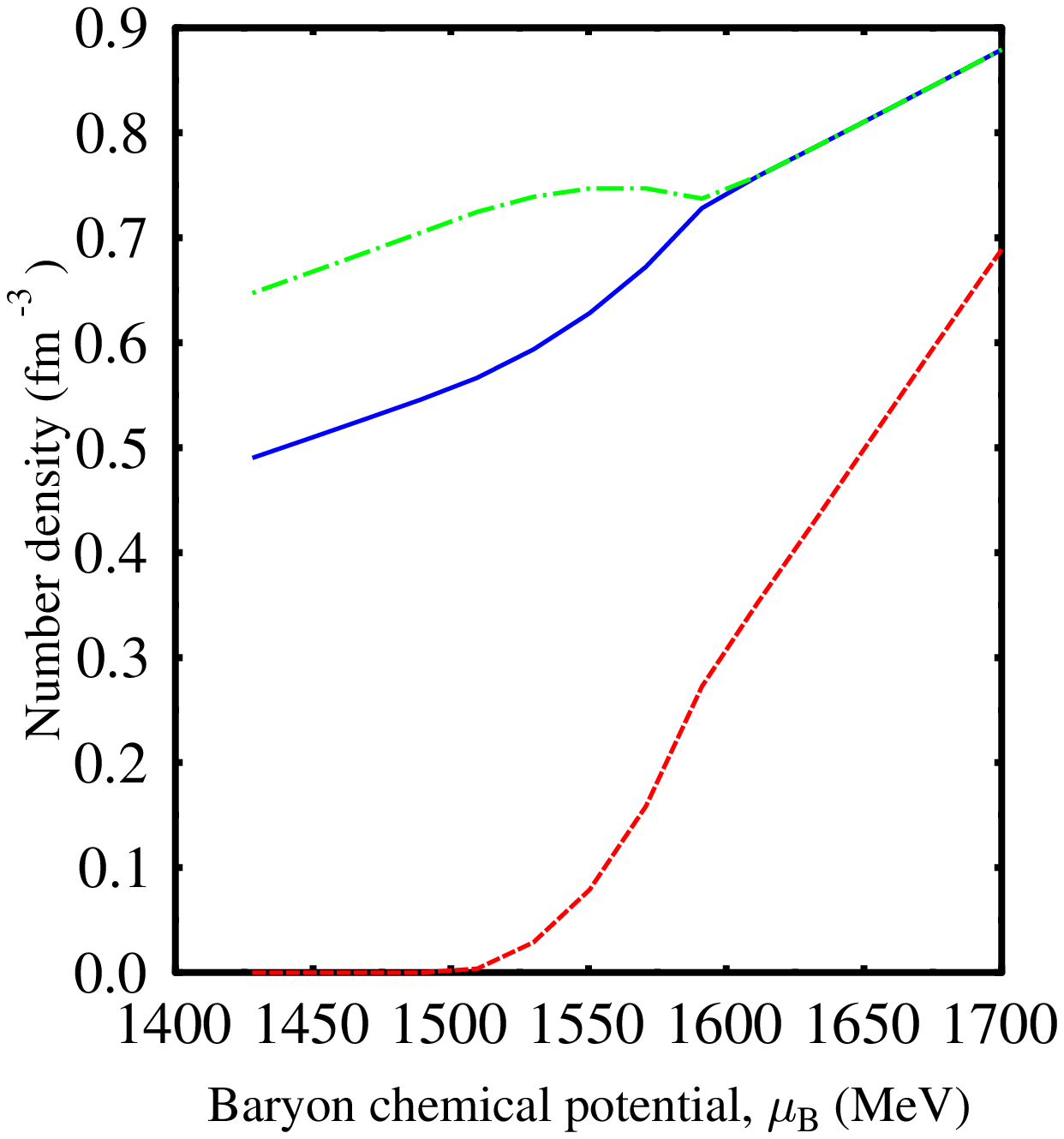}
\end{center}
\caption{\em Number densities  of u quarks (solid) and d quarks (dot-dashed) 
participating in superconducting phase. The density of s- quarks is also
plotted (dashed curve).}
\label{figdensc}
 \end{figure}

At high densities when chiral symmetry is restored for the light quarks, 
$\delta_\epsilon=0$ and $\delta=-\delta_\nu=-(\mu_E/2-(g^2/8)(\rho^u-\rho^d))$.
For $\delta_\nu < 0$, $\omega_1$, the quasi particle energy for u quarks is 
always positive and hence only the d quark contribution will be there in the
second term of the thermodynamic potential given in Eq.(\ref{omgud}).
Further, it is easy to show that such a contribution arises when
the magnitude of $\delta_\nu$,
which is half the difference of the chemical potentials of
the quarks which condense, is the same or greater than the
superconducting gap, $\Delta$.
When $|\delta_\nu|=\Delta$, the mode
$\omega_2=\sqrt{\Delta^2+(\bar\epsilon^2-\bar \nu^2)}-\delta$, becomes
gapless at the fermi sphere. For $|\delta_\nu| > \Delta$, 
the gapless modes occur
at momenta higher than the (average) fermi momenta.
In the present calculation this 
is the case for baryon chemical potential below  $1600$ MeV. The occurrence of
such {\em gapless superconducting modes} in neutral quark matter
was first emphasised in Ref.\cite{igor}. Because of this the 
number densities of $u$ and $d$ quarks participating in the superconducting
phase are not the same in this region. This is plotted in Fig. \ref{figdensc}.
It may be noted that even in the gapless
superconducting mode, the  density of strange quarks is 
small but nonzero.

\section{Summary}
                 We have analysed here in a current current point 
interaction model, the structure of vacuum in terms of quark antiquark 
as well as diquark pairs. The methodology uses an explicit variational 
construct of the trial state and is not based on a mean field calculations. 
Because of the point interaction structure we could solve for the gap functions
explicitly. The distribution functions are also determined variationally. 
Because of the vector interaction, the chemical potentials are interaction 
dependent and are calculated self consistently. We find that there is a small 
window in baryon chemical potential (about 80 MeV) when both the condensates
are nonzero.

To consider neutron star matter, we have imposed the condition 
of color and electric charge neutrality conditions through introduction of 
appropriate chemical potentials. It has been noted and emphasized earlier that
projecting out the color singlet state from color neutral state costs 
negligible free energy for large enough chunk of color neutral matter
\cite{krisch}.

The gap reduces when color and electric neutrality conditions are
imposed. The pressure with the gap is always higher than free quarks when
charge neutrality conditions are imposed. For slightly lower densities,
but large enough to be in the chiral symmetry restored phase, there appears to
be gapless modes available when the condensing quarks have a difference in
chemical potentials which is larger than twice the superconducting gap. 
In all these calculations
we have also included the effect of self consistently determined mass for the 
strange quark. In fact, in the gapless superconducting phase, 
the number densities of strange quarks is also nonzero.

We have focussed our attention here to the superconducting phase. 
The variational method adopted can be directly generalised 
to include color flavor locked 
phase and one can then make a free energy comparison
regarding possibility of which phase would be thermodynamicaly favourable
at what density. This will be particularly interesting for
cooling of neutron stars with a CFL core.
We have considered here homogeneous phase of matter. 
However, we can also consider mixed phases of matter with matter being neutral
on the average. Some of these problems are being investigated and will be 
reported elsewhere \cite{hmampp}.

\begin{acknowledgements}
The authors would like to thank J.C. Parikh, D. H. Rischke,
I. Shovkovy and M. Huang for many useful discussions.
One of the authors (AM) would like to acknowledge financial support from 
Bundesministerium f\"ur Bildung und Forschung (BMBF) and Deutsches
Elektronen Synchrotron (DESY),
and Institut fuer Theoretische Physik, Frankfurt University
for warm hospitality.
\end{acknowledgements}

\def \ltg{R.P. Feynman, Nucl. Phys. B 188, 479 (1981); 
K.G. Wilson, Phys. Rev. \zbf  D10, 2445 (1974); J.B. Kogut,
Rev. Mod. Phys. \zbf  51, 659 (1979); ibid  \zbf 55, 775 (1983);
M. Creutz, Phys. Rev. Lett. 45, 313 (1980); ibid Phys. Rev. D21, 2308
(1980); T. Celik, J. Engels and H. Satz, Phys. Lett. B129, 323 (1983)}

\def\berges {J. Berges, K. Rajagopal, {\NPB{538}{215}{1999}}.}
\def \svz {M.A. Shifman, A.I. Vainshtein and V.I. Zakharov,
Nucl. Phys. B147, 385, 448 and 519 (1979);
R.A. Bertlmann, Acta Physica Austriaca 53, 305 (1981)}

\def \spmbst {S.P. Misra, Phys. Rev. D35, 2607 (1987)}

\def \hmgrnv { H. Mishra, S.P. Misra and A. Mishra,
Int. J. Mod. Phys. A3, 2331 (1988)}

\def \snss {A. Mishra, H. Mishra, S.P. Misra
and S.N. Nayak, Phys. Lett 251B, 541 (1990)}

\def \amqcd { A. Mishra, H. Mishra, S.P. Misra and S.N. Nayak,
Pramana (J. of Phys.) 37, 59 (1991). }
\def\qcdtb{A. Mishra, H. Mishra, S.P. Misra 
and S.N. Nayak, Z.  Phys. C 57, 233 (1993); A. Mishra, H. Mishra
and S.P. Misra, Z. Phys. C 58, 405 (1993)}

\def \spmtlk {S.P. Misra, Talk on {\it `Phase transitions in quantum field
theory'} in the Symposium on Statistical Mechanics and Quantum field theory, 
Calcutta, January, 1992, hep-ph/9212287}

\def \hmnj {H. Mishra and S.P. Misra, 
{\PRD{48}{5376}{1993}.}}

\def \hmqcd {A. Mishra, H. Mishra, V. Sheel, S.P. Misra and P.K. Panda,
hep-ph/9404255 (1994)}

\def \amcrl {A. Mishra, H. Mishra and S.P. Misra, Z. Phys. C 57, 241 (1993)}

\def \higgs { S.P. Misra, in {\it Phenomenology in Standard Model and Beyond}, 
Proceedings of the Workshop on High Energy Physics Phenomenology, Bombay,
edited by D.P. Roy and P. Roy (World Scientific, Singapore, 1989), p.346;
A. Mishra, H. Mishra, S.P. Misra and S.N. Nayak, Phys. Rev. D44, 110 (1991)}

\def \nmtr {A. Mishra, 
H. Mishra and S.P. Misra, Int. J. Mod. Phys. A5, 3391 (1990); H. Mishra,
 S.P. Misra, P.K. Panda and B.K. Parida, Int. J. Mod. Phys. E 1, 405, (1992);
 {\it ibid}, E 2, 547 (1993); A. Mishra, P.K. Panda, S. Schrum, J. Reinhardt
and W. Greiner, to appear in Phys. Rev. C}

\def \dtrn {P.K. Panda, R. Sahu and S.P. Misra, 
Phys. Rev C45, 2079 (1992)}

\def \qcd {G. K. Savvidy, Phys. Lett. 71B, 133 (1977);
S. G. Matinyan and G. K. Savvidy, Nucl. Phys. B134, 539 (1978); N. K. Nielsen
and P. Olesen, Nucl.  Phys. B144, 376 (1978); T. H. Hansson, K. Johnson,
C. Peterson Phys. Rev. D26, 2069 (1982)}

\def \cornwal {J.M. Cornwall, Phys. Rev. D26, 1453 (1982)}
\def\aichlin {F. Gastineau, R. Nebauer and J. Aichelin,
{\PRC{65}{045204}{2002}}.}

\def \mndglv {J. E. Mandula and M. Ogilvie, Phys. Lett. 185B, 127 (1987)}

\def \schwinger {J. Schwinger, Phys. Rev. 125, 1043 (1962); ibid,
127, 324 (1962)}

\def \schutte {D. Schutte, Phys. Rev. D31, 810 (1985)}

\def \amspm {A. Mishra and S.P. Misra, 
{\ZPC{58}{325}{1993}}.}

\def \gft{ For gauge fields in general, see e.g. E.S. Abers and 
B.W. Lee, Phys. Rep. 9C, 1 (1973)}

\def \gribov {V.N. Gribov, Nucl. Phys. B139, 1 (1978)}

\def \spm78 {S.P. Misra, Phys. Rev. D18, 1661 (1978); {\it ibid}
D18, 1673 (1978)} 

\def \lopr {A. Le Youanc, L.  Oliver, S. Ono, O. Pene and J.C. Raynal, 
Phys. Rev. Lett. 54, 506 (1985)}

\def \spphi {S.P. Misra and S. Panda, Pramana (J. Phys.) 27, 523 (1986);
S.P. Misra, {\it Proceedings of the Second Asia-Pacific Physics Conference},
edited by S. Chandrasekhar (World Scientfic, 1987) p. 369}

\def\spmdif {S.P. Misra and L. Maharana, Phys. Rev. D18, 4103 (1978); 
    S.P. Misra, A.R. Panda and B.K. Parida, Phys. Rev. Lett. 45, 322 (1980);
    S.P. Misra, A.R. Panda and B.K. Parida, Phys. Rev. D22, 1574 (1980)}

\def \spmvdm {S.P. Misra and L. Maharana, Phys. Rev. D18, 4018 (1978);
     S.P. Misra, L. Maharana and A.R. Panda, Phys. Rev. D22, 2744 (1980);
     L. Maharana,  S.P. Misra and A.R. Panda, Phys. Rev. D26, 1175 (1982)}

\def\spmthr {K. Biswal and S.P. Misra, Phys. Rev. D26, 3020 (1982);
               S.P. Misra, Phys. Rev. D28, 1169 (1983)}

\def \spmstr { S.P. Misra, Phys. Rev. D21, 1231 (1980)} 

\def \spmjet {S.P. Misra, A.R. Panda and B.K. Parida, Phys. Rev Lett. 
45, 322 (1980); S.P. Misra and A.R. Panda, Phys. Rev. D21, 3094 (1980);
  S.P. Misra, A.R. Panda and B.K. Parida, Phys. Rev. D23, 742 (1981);
  {\it ibid} D25, 2925 (1982)}

\def \arpftm {L. Maharana, A. Nath and A.R. Panda, Mod. Phys. Lett. 7, 
2275 (1992)}

\def \van {R. Van Royen and V.F. Weisskopf, Nuov. Cim. 51A, 617 (1965)}

\def \rchpi {S.R. Amendolia {\it et al}, Nucl. Phys. B277, 168 (1986)}

\def \chrl{ Y. Nambu, {\PRL{4}{380}{1960}};
A. Amer, A. Le Yaouanc, L. Oliver, O. Pene and
J.C. Raynal,{\PRL{50}{87}{1983a}};{\em ibid}
{\PRD{28}{1530}{1983}};
M.G. Mitchard, A.C. Davis and A.J.
MAacfarlane, {\NPB{325}{470}{1989}};
B. Haeri and M.B. Haeri,{\PRD{43}{3732}{1991}};
V. Bernard,{\PRD{34}{1604}{1986}};
 S. Schramm and
W. Greiner, Int. J. Mod. Phys. \zbf E1, 73 (1992), 
J.R. Finger and J.E. Mandula, Nucl. Phys. \zbf B199, 168 (1982),
S.L. Adler and A.C. Davis, Nucl. Phys.\zbf  B244, 469 (1984),
S.P. Klevensky, Rev. Mod. Phys.\zbf  64, 649 (1992).}

\def \spmijp { S.P. Misra, Ind. J. Phys. 61B, 287 (1987)}

\def \feynman {R.P. Feynman and A.R. Hibbs, {\it Quantum mechanics and
path integrals}, McGraw Hill, New York (1965)}

\def \glstn{ J. Goldstone, Nuov. Cim. \zbf 19, 154 (1961);
J. Goldstone, A. Salam and S. Weinberg, Phys. Rev. \zbf  127,
965 (1962)}

\def \anderson {P.W. Anderson, Phys. Rev. \zbf {110}, 827 (1958)}

\def \nambu{ Y. Nambu, Phys. Rev. Lett. \zbf 4, 380 (1960)}

\def\donogh {J.F. Donoghue, E. Golowich and B.R. Holstein, {\it Dynamics
of the Standard Model}, Cambridge University Press (1992)}

\def\satz {T. Matsui and H. Satz, Phys. Lett. B178, 416 (1986)}

\def\cps {C. P. Singh, Phys. Rep. 236, 149 (1993)}

\def\prliop {A. Mishra, H. Mishra, S.P. Misra, P.K. Panda and Varun
Sheel, Int. J. of Mod. Phys. E 5, 93 (1996)}

\def\hmcor {V. Sheel, H. Mishra and J.C. Parikh, Phys. Lett. B382, 173
(1996); {\it biid}, to appear in Int. J. of Mod. Phys. E}
\def\cort { V. Sheel, H. Mishra and J.C. Parikh, Phys. ReV D59,034501 (1999);
{\it ibid}Prog. Theor. Phys. Suppl.,129,137, (1997).}

\def\surcor {E.V. Shuryak, Rev. Mod. Phys. 65, 1 (1993)} 

\def\stevenson {A.C. Mattingly and P.M. Stevenson, Phys. Rev. Lett. 69,
1320 (1992); Phys. Rev. D 49, 437 (1994)}

\def\mac {M. G. Mitchard, A. C. Davis and A. J. Macfarlane,
 Nucl. Phys. B 325, 470 (1989)} 
\def\tfd
 {H.~Umezawa, H.~Matsumoto and M.~Tachiki {\it Thermofield dynamics
and condensed states} (North Holland, Amsterdam, 1982) ;
P.A.~Henning, Phys.~Rep.253, 235 (1995).}
\def\amph4{Amruta Mishra and Hiranmaya Mishra,
{\JPG{23}{143}{1997}}.}

\def \neglecor{M.-C. Chu, J. M. Grandy, S. Huang and 
J. W. Negele, Phys. Rev. D48, 3340 (1993);
ibid, Phys. Rev. D49, 6039 (1994)}

\def\revdata {Particle Data Group, Phys. Rev. D 50, 1173 (1994)}

\def\sinp {S.P. Misra, Indian J. Phys., {\bf 70A}, 355 (1996)}
\def\npa{H. Mishra and J.C. Parikh, {\NPA{679}{597}{2001}.}}
\def\krisch {M. Alford and K. Rajagopal, JHEP 0206,031,(2002)}
\def\reddy {A.W. Steiner, S. Reddy and M. Prakash,
{\PRD{66}{094007}{2002}.}}

\def\bryman {D.A. Bryman, P. Deppomier and C. Le Roy, Phys. Rep. 88,
151 (1982)}

\def\thooft {G. 't Hooft, Phys. Rev. D 14, 3432 (1976); D 18, 2199 (1978);
S. Klimt, M. Lutz, U. Vogl and W. Weise, Nucl. Phys. A 516, 429 (1990)}
\def\alkz { R. Alkofer, P. A. Amundsen and K. Langfeld, Z. Phys. C 42,
199(1989), A.C. Davis and A.M. Matheson, Nucl. Phys. B246, 203 (1984).}
\def\sarah {T.M. Schwartz, S.P. Klevansky, G. Papp,
{\PRC{60}{055205}{1999}}.}
\def\wil{M. Alford, K.Rajagopal, F. Wilczek, {\PLB{422}{247}{1998}};
{\it{ibid}}{\NPB{537}{443}{1999}}.}
\def\sursc{R.Rapp, T.Schaefer, E. Shuryak and M. Velkovsky,
{\PRL{81}{53}{1998}};{\it ibid}{\AP{280}{35}{2000}}.}
\def\pisarski{
D. Bailin and A. Love, {\PR{107}{325}{1984}},
D. Son, {\PRD{59}{094019}{1999}}; 
T. Schaefer and F. Wilczek, {\PRD{60}{114033}{1999}};
D. Rischke and R. Pisarski, {\PRD{61}{051501}{2000}}, 
D. K. Hong, V. A. Miransky, 
I. A. Shovkovy, L.C. Wiejewardhana, {\PRD{61}{056001}{2000}}.}
\def\leblac {M. Le Bellac, {\it Thermal Field Theory}(Cambridge, Cambridge University
Press, 1996).}
\def\bcs{A.L. Fetter and J.D. Walecka, {\it Quantum Theory of Many
particle Systems} (McGraw-Hill, New York, 1971).}
\def\alexander{Aleksander Kocic, Phys. Rev. D33, 1785,(1986).}
\def\bubmix{F. Neumann, M. Buballa and M. Oertel,
{\NPA{714}{481}{2003}.}}
\def\kunihiro{M. Kitazawa, T. Koide, T. Kunihiro, Y. Nemeto,
{\PTP{108}{929}{2002}.}}
\def\igor{Igor Shovkovy, Mei Huang, hep-ph/0302142; I. Shovkovy,
M. Hanauske and M. Huang, 
{\PRD{67}{103004}{2003}}.}
\def\abrikosov{A.A. Abrikosov, L.P. Gorkov, Zh. Eskp. Teor.39, 1781,
1960}
\def\krischprl{M.G. Alford, J. Berges and K. Rajagopal,
 {\PRL{84}{598}{2000}.}}
\def\hmampp{A. Mishra and H.Mishra, in preparation}
\def\blaschke{D. Blaschke, M.K. Volkov and V.L. Yudichev,
{\EPJA{17}{103}{2003}}.}
\def\mei{M. Huang, P. Zhuang, W. Chao,
{\PRD{65}{076012}{2002}}}
\def\bubnp{M. Buballa, M. Oertel,
{\NPA{703}{770}{2002}};
P. Rehberg, S.P. Klevansky and J. Huefner,
{\PRC{53}{410}{1996}.}}

\end{document}